\documentclass[11pt]{article}

 \usepackage{amsmath}
 \usepackage{tabularx}
 \usepackage{array}
 \usepackage{booktabs}
 \usepackage{graphicx}
 \usepackage{longtable}
 
 \usepackage{amsfonts}
 \usepackage{float}
\usepackage{hyperref}
\usepackage{bm, geometry, cite, amssymb}
\begin{document}
 
 \title{\LARGE \bf {A Geometric Framework for CPT Violation in Neutral Meson Mixing Using Biorthogonal Bargmann Invariants}}

 \author{
 	Swarup Sangiri \thanks{swarup.phys@gmail.com} \\
   \textit{Physics Department,
	Indian Institute of Technology Kharagpur}\\ \textit{Kharagpur, 721302, India}
}
\date{}
\maketitle

\begin{abstract}
We develop a geometric framework for characterizing CPT violation in neutral meson systems using Bargmann invariants formulated within a biorthogonal description of the non-Hermitian effective Hamiltonian governing neutral meson mixing. Interpreting CPT violation as a relative geometric deformation of the heavy- and light-state mixing directions in projective flavor space, we construct a fourth-order Bargmann invariant together with its CP-conjugate counterpart involving the physical mass eigenstates and experimentally accessible decay channels. From the phase of a rephasing-invariant product of these invariants, we define a geometric observable that isolates the CPT-violating contribution. The resulting formalism identifies the channel dependence of the geometric response and yields a selection criterion for decay-mode combinations exhibiting linear sensitivity to CPT violation. We further relate the geometric deformation to the Lorentz-violating coefficients of the Standard-Model Extension, showing that the resulting observable inherits the characteristic sidereal modulation of the SME framework. The present work provides a complementary geometric perspective on CPT violation in neutral meson mixing and establishes a foundation for future phenomenological studies of geometric signatures of CPT and Lorentz violation.
\end{abstract}
\newpage
\section{Introduction}
Neutral meson systems provide one of the most sensitive laboratories for testing the fundamental discrete symmetries of nature \cite{BigiSanda2009, Branco1999}. The phenomenon of particle--antiparticle mixing, together with the interference between
mixing and decay, has enabled precision investigations of CP, T, and CPT symmetries over several decades and continues to play a central role in searches for physics beyond the Standard Model \cite{Christenson1964, BigiSanda1981, ColladayKostelecky1997, PDG2024}. Owing to their unstable nature, these systems are described by an effective non-Hermitian Hamiltonian whose eigenvalues determine the physical masses and decay widths, while the corresponding eigenstates encode the flavor-mixing dynamics \cite{LeeOehmeYang1957}. Consequently, neutral mesons provide a unique setting in which symmetry violation, quantum interference, and non-Hermitian evolution can be studied within a common theoretical framework \cite{Domenico2012, Ohlsson2021}.

Among the possible extensions of the Standard Model, violations of CPT symmetry are of particular interest because the CPT theorem guarantees CPT invariance within local relativistic quantum field theory \cite{Luderes1957, Pauli1955, Lehnert2016}. Any experimentally established CPT violation would therefore constitute unambiguous evidence for new physics. Neutral meson oscillations have
accordingly become one of the primary systems for investigating CPT- and Lorentz-violating effects, both through conventional effective-Hamiltonian analyses and within the framework of the Standard-Model Extension (SME), which provides a systematic effective field-theoretic description of such phenomena \cite{ColladayKostelecky1998, Kostelecky1999, Greenberg2002}.

In parallel with these developments, geometric concepts have acquired an important role in quantum theory. Since the discovery of the geometric phase, it has become evident that the evolution of a quantum system contains information determined not only by its dynamics but also by the geometry of the underlying projective Hilbert space \cite{Pancharatnam1956, Berry1984, AharonovAnandan1987, SamuelBhandari1988, GarrisonWright1988}. Within this framework, Bargmann invariants (BIs) provide a particularly natural geometric description by constructing rephasing-invariant quantities from closed sequences of quantum states, thereby encoding geometric information independently of arbitrary basis choices \cite{Bargmann1964, Rabei1999, Mukunda2003}. Such invariants have found applications in quantum information, topological phases, non-Hermitian quantum mechanics, and various interference phenomena \cite{MukundaSimon1993, Pratapsi2025, Zhang2026}.

Geometric phases have also attracted considerable attention in neutral meson physics, where they have been employed to investigate flavor oscillations, CP violation, and the geometric properties of unstable quantum systems. More recently, geometric formulations based on BIs have been developed for neutral meson mixing and correlated meson systems, establishing a rephasing-invariant description of the underlying geometric structure and its connection with CP-violating observables \cite{SangiriSarkar2023, Sangiri2026, Sangiri22026}. These developments naturally motivate the investigation of whether a similar rephasing-invariant geometric framework can be extended to CPT violation. In the present work, we investigate this question by constructing the corresponding geometric framework within a biorthogonal description of the effective Hamiltonian governing neutral meson mixing. In particular, it is of interest to examine whether BIs can provide a geometric characterization of CPT-violating effects and their connection to the underlying Lorentz- and CPT-violating dynamics.

Despite the extensive phenomenological studies of CPT violation in neutral mesons, most analyses have been formulated in terms of effective-Hamiltonian parameters, decay-rate asymmetries, or the Lorentz-violating coefficients of the Standard-Model Extension (SME), all of which provide well-established frameworks for describing CPT-violating effects \cite{Kostelecky2001, KosteleckyRoberts2001, Bernabeu2001}. In contrast, BIs furnish a natural rephasing-invariant geometric framework for characterizing quantum-state evolution. These considerations motivate the investigation of whether a corresponding BI formulation can be developed for CPT-violating neutral meson systems within the biorthogonal framework adopted in this work. Such an approach offers the possibility of relating the deformation of the neutral-meson mixing structure induced by CPT violation to a rephasing-invariant geometric observable.

The present work addresses this problem by developing a geometric framework for CPT violation based on BIs in non-Hermitian neutral meson systems. Employing a biorthogonal formulation of the effective Hamiltonian, we construct a fourth-order BI together with its CP-conjugate counterpart using the heavy- and light-meson eigenstates and experimentally accessible decay channels. From these invariants, we define a geometric observable that isolates the CPT-induced contribution and provides a natural geometric interpretation of the deformation of the mixing structure. The resulting formalism naturally leads to a channel-selection criterion governing the sensitivity of different decay modes, establishes a connection with the Lorentz-violating coefficients of the Standard-Model Extension, and demonstrates how the resulting geometric observable inherits the characteristic sidereal modulation associated with the Earth's rotation.

Rather than introducing an alternative phenomenological parametrization of CPT violation, the present work offers a complementary geometric viewpoint in which CPT-violating effects are characterized through a rephasing-invariant geometric observable constructed from Bargmann invariants. This formulation unifies the geometric description of neutral-meson mixing with the established phenomenology of Lorentz and CPT violation, while naturally incorporating the channel dependence of the geometric response and its connection to the Standard-Model Extension. It thereby provides a systematic foundation for future theoretical and phenomenological investigations of geometric signatures of CPT and Lorentz violation in precision neutral-meson experiments.

\section{Neutral Meson Mixing in the Presence of CPT Violation}
The geometric characterization of CPT violation developed in this work relies on the effective Hamiltonian description of neutral meson mixing \cite{LeeOehmeYang1957, Kostelecky2001, BigiSanda2009}. In this framework, departures from CPT invariance manifest themselves through an asymmetry between the diagonal elements of the non-Hermitian mixing Hamiltonian, leading to distinct heavy- and light-state mixing structures. In this section, we introduce the effective-Hamiltonian description and establish the notation and conventions used throughout the remainder of the paper. The time evolution of a neutral meson--antimeson system,
\(\left\{|P^0\rangle,|\overline {P^0}\rangle\right\}\), where
\(P=K,D,B_d,B_s\), is governed by the Schr\"odinger equation
\begin{equation}
i\frac{d}{dt}
\begin{pmatrix}
|P^0(t)\rangle \\
|\overline {P^0}(t)\rangle
\end{pmatrix}
=
H
\begin{pmatrix}
|P^0(t)\rangle \\
|\overline {P^0}(t)\rangle
\end{pmatrix},
\label{eq:schrodinger}
\end{equation}
where the effective Hamiltonian is conventionally written as $H=M-\frac{i}{2}\Gamma$. Here, \(M\) and \(\Gamma\) are \(2\times2\) Hermitian matrices describing the dispersive and absorptive components of the mixing dynamics, respectively. In the flavor basis
\(\{|P^0\rangle,|\overline {P^0}\rangle\}\), the effective Hamiltonian takes
the form
\begin{equation}
H=
\begin{pmatrix}
H_{11} & H_{12} \\
H_{21} & H_{22}
\end{pmatrix}.
\label{eq:H_matrix}
\end{equation}
Violation of CPT symmetry manifests itself through a difference in the diagonal matrix elements, $H_{11}\neq H_{22}$, whereas flavor mixing is governed by the off-diagonal elements $H_{12}$ and $H_{21}$. Violation of T symmetry is associated with an asymmetry between these elements, while CP violation may arise through the combined effects of CPT- and T-violating contributions.

The heavy (\(H\)) and light (\(L\)) mass eigenstates are defined by
\begin{equation}
H|P_{H,L}\rangle
=
\lambda_{H,L}|P_{H,L}\rangle,
\label{eq:eigenvalue_equation}
\end{equation}
with complex eigenvalues
\begin{equation}
\lambda_{H,L}
=
m_{H,L}
-\frac{i}{2}\Gamma_{H,L}.
\label{eq:eigenvalues}
\end{equation}
The eigenvalue difference is given by
\begin{equation}
\Delta\lambda
\equiv
\lambda_H-\lambda_L
=
\Delta m
-\frac{i}{2}\Delta\Gamma,
\label{eq:delta_lambda}
\end{equation}
where $\Delta m=m_H-m_L$, and $\Delta\Gamma=\Gamma_H-\Gamma_L$. Throughout this work, we adopt the convention
\begin{subequations}
    \begin{align}
     |P_H\rangle &=p_H|P^0\rangle+q_H|\overline {P^0}\rangle \label{eq:PH_definition}\\
     |P_L\rangle &=p_L|P^0\rangle-q_L|\overline {P^0}\rangle \label{eq:PL_definition}   
    \end{align}
\end{subequations}

where \(p_{H,L}\) and \(q_{H,L}\) are complex mixing coefficients specifying the flavor composition of the heavy and light mass eigenstates.

Solving Eq.~\eqref{eq:eigenvalue_equation} yields $(H_{11}-\lambda_H)p_H+H_{12}q_H=0$ and $(H_{11}-\lambda_L)p_L-H_{12}q_L=0$, which gives the mixing ratios 
\begin{equation}
\frac{q_H}{p_H}
=
-\frac{H_{11}-\lambda_H}{H_{12}},
\qquad
\frac{q_L}{p_L}
=
\frac{H_{11}-\lambda_L}{H_{12}}.
\label{eq:qp_exact}
\end{equation}
The eigenvalues of the effective Hamiltonian are obtained from the
characteristic equation,
\begin{equation}
\lambda_{H,L}
=
\frac{H_{11}+H_{22}}{2}
\pm
\frac{1}{2}
\sqrt{\left(H_{11}-H_{22}\right)^2
+
4H_{12}H_{21}},
\label{eq:lambda_exact}
\end{equation}
from which the eigenvalue difference satisfies
\begin{equation}
(\Delta\lambda)^2
=
\left(H_{11}-H_{22}\right)^2
+
4H_{12}H_{21}.
\label{eq:delta_lambda_exact}
\end{equation}
To characterize CPT violation, we introduce the dimensionless parameter
\begin{equation}
z
\equiv
\frac{H_{11}-H_{22}}{\Delta\lambda}.
\label{eq:z_definition}
\end{equation}
The parameter \(z\) therefore measures the departure of the heavy- and light-state flavor compositions from the CPT-conserving limit and will serve as the fundamental small parameter governing the geometric deformation discussed below.
It is convenient to introduce the quantity
\begin{equation}
r\equiv
\sqrt{\frac{H_{21}}{H_{12}}},
\label{eq:r_definition}
\end{equation}
which characterizes the relative magnitude and phase of the off-diagonal mixing amplitudes. The branch of the square root is chosen such that the phase convention of the mass eigenstates is preserved throughout the analysis. Using the exact eigenvalues of the effective Hamiltonian in Eq.~(\ref{eq:lambda_exact}), together with the definition in Eq.~(\ref{eq:z_definition}),
one finds
\begin{equation}
H_{11}-\lambda_H
=
-\frac{\Delta\lambda}{2}(1-z),
\qquad
H_{11}-\lambda_L
=
\frac{\Delta\lambda}{2}(1+z).
\end{equation}
Furthermore, from Eq.~(\ref{eq:delta_lambda_exact}), it follows that
\begin{equation}
\Delta\lambda
=
\frac{2\sqrt{H_{12}H_{21}}}
{\sqrt{1-z^2}}.
\end{equation}
Substituting these relations into Eq.~\eqref{eq:qp_exact} yields
\begin{equation}
\frac{q_H}{p_H}
=
r
\sqrt{\frac{1-z}{1+z}},
\label{eq:qHpH}
\end{equation}
and
\begin{equation}
\frac{q_L}{p_L}
=
r
\sqrt{\frac{1+z}{1-z}}.
\label{eq:qLpL}
\end{equation}

In the CPT-conserving limit, \(z=0\), one recovers the familiar
relation
\begin{equation}
\frac{q_H}{p_H}
=
\frac{q_L}{p_L}
=
r.
\end{equation}

Equations~\eqref{eq:qHpH} and \eqref{eq:qLpL} admit a natural
geometric interpretation of CPT violation. The quantities
\(q_H/p_H\) and \(q_L/p_L\) determine the orientation of the heavy- and
light-mass eigenstates in the projective flavor space. CPT conservation
implies that these directions coincide, whereas CPT violation induces a
relative deformation between them:
\begin{equation}
\frac{q_H}{p_H}
\neq
\frac{q_L}{p_L},
\label{eq:geometric_interpretation}
\end{equation}
when $z\neq0$. This observation motivates the geometric construction developed in the following sections. In particular, the mismatch between the heavy- and light-state mixing directions is used to construct a rephasing-invariant geometric observable sensitive to CPT violation.

Within the framework of the Standard-Model Extension (SME), the CPT-violating parameter $z$ admits a direct interpretation in terms of the Lorentz- and CPT-violating coefficients governing neutral-meson mixing \cite{Kostelecky1999,Kostelecky2001}. This correspondence provides the dynamical origin of the geometric deformation introduced above and implies a characteristic dependence on the meson four-velocity, giving rise to sidereal variations in the presence of Lorentz violation. The explicit connection with the SME coefficients and its implications for the geometric observable developed in this work will be established in a later section.

\section{Biorthogonal Formalism for Neutral Meson Systems}
The non-Hermitian effective Hamiltonian introduced in the previous section admits a generalized spectral description in which the left and right eigenvectors are treated independently \cite{Scholtz1992, Moiseyev2011, Brody2014}. In the present work, we adopt the corresponding biorthogonal formalism as the mathematical framework for constructing the geometric quantities developed in the following sections.

The right eigenstates of the effective Hamiltonian are defined by
\begin{equation}
H|P_i\rangle=\lambda_i|P_i\rangle,
\qquad
i=H,L,
\label{eq:right_eigenstates}
\end{equation}
where the complex eigenvalues \(\lambda_i\) were introduced in Eq.~\eqref{eq:eigenvalues}. Within the biorthogonal formulation adopted in the present work, the corresponding left eigenstates are defined through
\begin{equation}
\langle\widetilde{P}_i|H
=
\lambda_i\langle\widetilde{P}_i|,
\qquad
i=H,L.
\label{eq:left_eigenstates}
\end{equation}

For a diagonalizable effective Hamiltonian, the left and right eigenstates form a biorthogonal basis obeying
\begin{equation}
\langle\widetilde{P}_i|P_j\rangle
=
\delta_{ij},
\label{eq:biorthogonality}
\end{equation}
together with the completeness relation
\begin{equation}
\sum_{i=H,L}
|P_i\rangle\langle\widetilde{P}_i|
=
\mathbb{I}.
\label{eq:completeness}
\end{equation}

Using the conventions introduced in Eqs.~\eqref{eq:PH_definition} and \eqref{eq:PL_definition}, the left eigenstates may be written in the flavor basis as
\begin{subequations}
    \begin{align}
     \langle\widetilde{P}_H|&=\alpha_H\langle P^0|+\beta_H\langle\overline {P^0}|,
     \label{eq:leftH_ansatz}
     \\
     \langle\widetilde{P}_L|&=\alpha_L\langle P^0|+\beta_L\langle\overline {P^0}|.\label{eq:leftL_ansatz}
    \end{align}
\end{subequations}

The coefficients \(\alpha_i\) and \(\beta_i\) are determined by imposing the biorthogonality conditions in Eq.~\eqref{eq:biorthogonality}. A straightforward calculation yields
\begin{subequations}
    \begin{align}
     \langle\widetilde{P}_H|&=\frac{q_L\langle P^0|+p_L\langle\overline {P^0}|}{p_Hq_L+p_Lq_H},\label{eq:leftH} \\
     \langle\widetilde{P}_L|&=\frac{q_H\langle P^0|-p_H\langle\overline {P^0}|}{p_Hq_L+p_Lq_H}.\label{eq:leftL}
    \end{align}
\end{subequations}
These expressions satisfy Eq.~\eqref{eq:left_eigenstates}, together with the biorthogonality conditions $\langle\widetilde{P}_H|P_H\rangle=\langle\widetilde{P}_L|P_L\rangle=1$, and $\langle\widetilde{P}_H|P_L\rangle=\langle\widetilde{P}_L|P_H\rangle=0$. The normalization is well defined provided $p_Hq_L+p_Lq_H\neq 0$, corresponding to the generic non-degenerate case. The biorthogonal basis permits a spectral decomposition of the effective Hamiltonian of the form
\begin{equation}
H
=
\lambda_H
|P_H\rangle\langle\widetilde{P}_H|
+
\lambda_L
|P_L\rangle\langle\widetilde{P}_L|.
\label{eq:spectral_decomposition}
\end{equation}

To connect the mixing formalism with physical decay channels, we introduce a set of state-overlap amplitudes associated with the decay-channel rays. In conventional neutral-meson phenomenology, decay processes are described by the weak transition matrix elements \cite{Nir2005},
\begin{equation}
A_f^{\rm (phys)}
=
\langle f|T|P^0\rangle,
\qquad
\bar A_f^{\rm (phys)}
=
\langle f|T|\overline {P^0}\rangle,
\end{equation}
where \(T\) denotes the decay transition operator. In the present geometric framework, however, BIs are constructed from chains of quantum-state overlaps rather than transition matrix elements. We therefore associate each experimentally accessible decay channel with a corresponding ray \(|f\rangle\) and define the geometric overlap amplitudes
\begin{equation}
A_f
\equiv
\langle f|P^0\rangle,
\qquad
\bar A_f
\equiv
\langle f|\overline {P^0}\rangle,
\label{eq:decay_amplitudes}
\end{equation}
for a generic final state \(|f\rangle\). When the decay-channel rays are chosen to represent the corresponding physical decay channels, these overlap amplitudes are proportional to the conventional weak decay matrix elements, with the overall proportionality factors canceling in the rephasing-invariant quantities considered below. Thus, \(A_f\) and \(\bar A_f\) should be regarded as geometric overlap amplitudes characterizing the projection of the flavor states onto the state \(|f\rangle\), where \(|f\rangle\) denotes the effective projective representative of the corresponding physical decay channel entering the Bargmann construction rather than the asymptotic multiparticle state itself.

The quantities introduced in this section provide the ingredients required for the geometric construction developed in the following section. In particular, the biorthogonal pairing between left and right eigenstates forms the basis of the BI construction adopted in the present work.

\section{Construction of Bargmann Invariants in the Biorthogonal Formalism}

Having established the biorthogonal description of neutral meson mixing, we now construct the geometric quantities that form the basis of our analysis. Bargmann invariants (BIs) provide a geometric characterization of quantum evolution by encoding the relative phase accumulated along closed sequences of quantum states in projective Hilbert space \cite{Bargmann1964, Mukunda2003}. In the present work, we formulate the corresponding construction within the biorthogonal framework introduced in the previous section, which serves as the mathematical setting adopted throughout our analysis.

For a sequence of quantum states
\(\{|\psi_1\rangle,|\psi_2\rangle,\ldots,|\psi_n\rangle\}\),
the \(n\)-th order BI is defined as \cite{Bargmann1964, Mukunda2003}
\begin{equation}
\Delta_n
=
(\psi_1,\psi_2)
(\psi_2,\psi_3)
\cdots
(\psi_n,\psi_1),
\label{eq:Bargmann_general}
\end{equation}
where $(\psi_i,\psi_j)=\langle\psi_i|\psi_j\rangle$. Under independent local phase transformations,
\(|\psi_i\rangle\rightarrow e^{i\alpha_i}|\psi_i\rangle\),
the phases associated with adjacent overlaps cancel pairwise, rendering
\(\Delta_n\) invariant under arbitrary rephasings. Consequently, the associated geometric phase $\gamma_n=\arg\Delta_n$, depends only on the geometry of the closed path in projective state space.

Adopting the biorthogonal formulation introduced above, we define the corresponding BI by replacing the ordinary inner products with the associated biorthogonal overlaps between left and right states:
\begin{equation}
\Delta_n
=
(\widetilde{\psi}_1,\psi_2)
(\widetilde{\psi}_2,\psi_3)
\cdots
(\widetilde{\psi}_n,\psi_1),
\label{eq:biorthogonal_Bargmann_general}
\end{equation}
which remains invariant under the simultaneous transformations
\begin{equation}
|\psi_i\rangle
\rightarrow
e^{i\alpha_i}|\psi_i\rangle,
\qquad
\langle\widetilde{\psi}_i|
\rightarrow
e^{-i\alpha_i}
\langle\widetilde{\psi}_i|.
\label{eq:biorthogonal_rephasing}
\end{equation}
The quantity \(\Delta_n\) therefore preserves the rephasing invariance of the conventional BI while defining the geometric quantity employed in the present biorthogonal formulation.

\begin{figure}[H]
\centering
\includegraphics[width=0.45\textwidth]{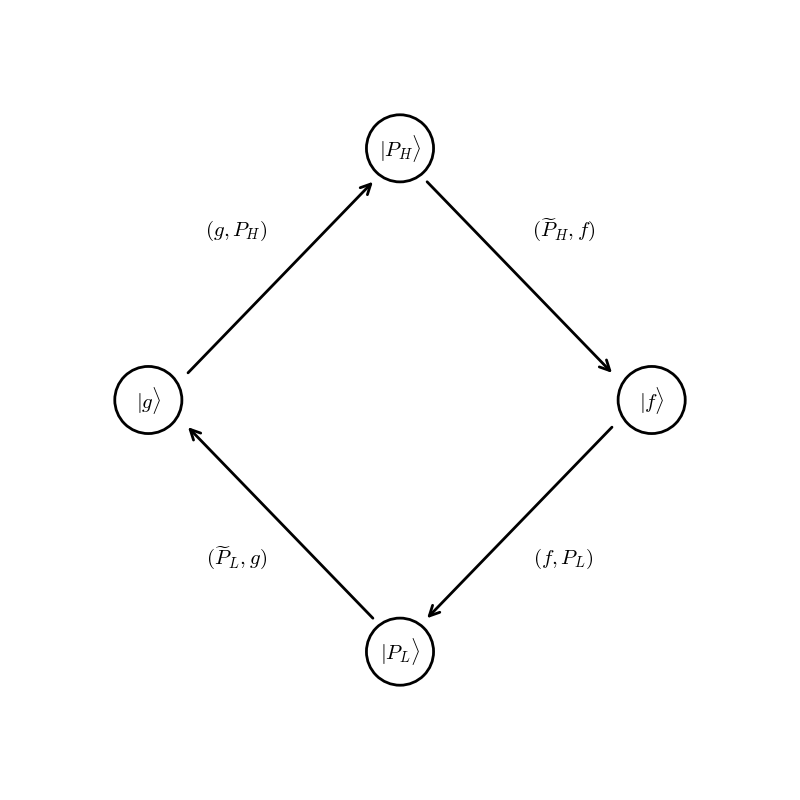}
\caption{Schematic representation of the fourth-order Bargmann invariant $\Delta_4^{(f,g)}$ as a closed cycle of biorthogonal overlaps connecting the heavy and light neutral-meson eigenstates through the decay-channel rays $|f\rangle$ and $|g\rangle$. The tilded states denote the left eigenstates of the non-Hermitian effective Hamiltonian.}
\label{fig:BI_fg}
\end{figure}

Specializing the above construction to neutral meson systems, we consider the minimal closed sequence involving the heavy and light mass eigenstates together with two experimentally distinguishable decay channels \(f\) and \(g\). Accordingly, the fourth-order BI takes the form
\begin{equation}
\Delta_4^{(f,g)}
=
(\widetilde{P}_H,f)
( f,P_L)
(\widetilde{P}_L,g)
(g,P_H).
\label{eq:Delta4_definition}
\end{equation}
The geometric structure of the fourth-order BI is illustrated schematically in Fig.~\ref{fig:BI_fg}. The closed cycle of biorthogonal overlaps provides a geometric representation of the invariant. Second-order invariants reduce to positive-definite intensities and therefore carry no geometric phase, while the natural third-order loops involving both heavy- and light-mass eigenstates collapse because
$(\widetilde P_H,P_L)=0$. The fourth-order construction of Eq.~(\ref{eq:Delta4_definition}) is therefore the lowest-order invariant capable of encoding nontrivial geometric information. Since the decay channels are asymptotic physical states belonging to an ordinary Hilbert space, only the unstable meson eigenstates require biorthogonal treatment. The final states \(f\) and \(g\) therefore enter through conventional Hermitian inner products. Because they are not eigenstates of the effective non-Hermitian Hamiltonian, no biorthogonal counterpart is required.

This invariant describes the closed sequence $P_H\rightarrow f\rightarrow P_L\rightarrow g \rightarrow P_H$, and defines a geometric phase through
\begin{equation}
\gamma_{fg}
=
\arg\Delta_4^{(f,g)}.
\label{eq:gamma_fg}
\end{equation}
Because $\Delta_4^{(f,g)}$ is invariant under independent flavor rephasings, the associated phase $\gamma_{fg}$ is likewise rephasing invariant.

Using Eqs.~\eqref{eq:leftH}, \eqref{eq:leftL}, \eqref{eq:PH_definition}, \eqref{eq:PL_definition}, and the decay amplitudes defined in Eq.~\eqref{eq:decay_amplitudes}, we obtain
\begin{subequations}
    \begin{align}
     (\widetilde{P}_H,f)
&=
\frac{
q_LA_f^{*}
+
p_L\bar{A}_f^{*}
}{
p_Hq_L+p_Lq_H
},\\
(f,P_L)
&=
p_LA_f-q_L\bar{A}_f,\\
(\widetilde{P}_L,g)
&=
\frac{
q_HA_g^{*}
-
p_H\bar{A}_g^{*}
}{
p_Hq_L+p_Lq_H
},\\
(g,P_H)
&=
p_HA_g+q_H\bar{A}_g.
    \end{align}
\end{subequations}

Substituting these expressions into Eq.~\eqref{eq:Delta4_definition}, the fourth-order invariant assumes the explicit form
\begin{align}
\Delta_4^{(f,g)}
=
\frac{
\left(q_LA_f^{*}+p_L\bar{A}_f^{*}\right)
\left(p_LA_f-q_L\bar{A}_f\right)
\left(q_HA_g^{*}-p_H\bar{A}_g^{*}\right)
\left(p_HA_g+q_H\bar{A}_g\right)
}{
\left(p_Hq_L+p_Lq_H\right)^2
}.
\label{eq:Delta4_explicit}
\end{align}

Under the flavor redefinitions $|P^0\rangle\rightarrow
e^{i\alpha}|P^0\rangle$, $|\overline{P^0}\rangle
\rightarrow e^{-i\alpha}|\overline{P^0}\rangle$, the mixing coefficients and decay amplitudes transform as
\begin{equation}
p_{H,L}\rightarrow e^{-i\alpha}p_{H,L},
\qquad
q_{H,L}\rightarrow e^{i\alpha}q_{H,L},
\end{equation}
\begin{equation}
A_{f,g}\rightarrow e^{i\alpha}A_{f,g},
\qquad
\bar{A}_{f,g}\rightarrow e^{-i\alpha}\bar{A}_{f,g}.
\end{equation}
It follows directly from Eq.~\eqref{eq:Delta4_explicit} that all phase factors cancel identically, rendering $\Delta_4^{(f,g)}$ manifestly rephasing invariant.

The fourth-order invariant defined in Eq.~\eqref{eq:Delta4_definition} therefore provides a rephasing-invariant characterization of the geometric structure associated with neutral meson mixing and decay. As we show in the following section, the combination of this invariant with its CP-conjugate counterpart enables the construction of a geometric probe of Lorentz- and CPT-violating effects.

\section{Geometric Characterization of CPT Violation}
The fourth-order BI constructed in the previous section provides a rephasing-invariant geometric characterization of the interplay between neutral meson mixing and decay. We now show how CPT violation enters this construction and derive a geometric observable that isolates the leading geometric contribution associated with the CPT-violating mixing parameter \(z\).

To identify the geometric signature of the deformation associated with the splitting of the heavy- and light-state directions in projective flavor space, we introduce the CP-conjugate counterpart of the invariant defined in Eq.~\eqref{eq:Delta4_definition}. Let \(|\bar f\rangle\) and \(|\bar g\rangle\) denote the CP-conjugate decay channels associated with \(|f\rangle\) and \(|g\rangle\), respectively. We further denote by $|\bar P_H\rangle$ and $|\bar P_L\rangle$ the states obtained from \(|P_H\rangle\) and \(|P_L\rangle\) under the CP transformation acting on the flavor basis, $|P^0\rangle
\overset{\mathrm{CP}}{\longleftrightarrow}
|\overline {P^0}\rangle$.

The corresponding fourth-order invariant is
\begin{equation}
\bar{\Delta}_4^{(\bar f,\bar g)}
=
\langle\widetilde{\bar P}_H|\bar f\rangle
\langle\bar f|\bar P_L\rangle
\langle\widetilde{\bar P}_L|\bar g\rangle
\langle\bar g|\bar P_H\rangle.
\label{eq:Delta4_CP}
\end{equation}

In the flavor basis, the matrix representation of the effective Hamiltonian transforms under CP through the interchange of particle and antiparticle components, implying
\begin{equation}
H_{11}
\overset{\mathrm{CP}}{\longleftrightarrow}
H_{22},
\qquad
H_{12}
\overset{\mathrm{CP}}{\longleftrightarrow}
H_{21}.
\label{eq:H_CP}
\end{equation}
Since CP changes the basis representation but does not alter the spectrum of the effective Hamiltonian, the eigenvalue difference $\Delta\lambda=\lambda_H-\lambda_L$ remains unchanged. Consequently,
\begin{equation}
z
=
\frac{H_{11}-H_{22}}{\Delta\lambda}
\overset{\mathrm{CP}}{\longrightarrow}
-z.
\label{eq:z_CP}
\end{equation}

Equation~\eqref{eq:z_CP} shows that the geometric deformation associated with CPT violation changes sign under CP conjugation. This observation motivates the construction of a rephasing-invariant geometric observable sensitive to CPT violation.

We therefore consider the product
\begin{equation}
\mathcal I_{\mathrm{CPT}}^{(f,g)}
=
\Delta_4^{(f,g)}
\,\,
\bar{\Delta}_4^{(\bar f,\bar g)},
\label{eq:I_CPT}
\end{equation}
and define the associated geometric observable
\begin{equation}
\mathcal G_{\mathrm{CPT}}^{(f,g)}
=
\arg\!\left(
\mathcal I_{\mathrm{CPT}}^{(f,g)}
\right).
\label{eq:G_CPT_definition}
\end{equation}

In the CPT-conserving limit ($z=0$), the heavy- and light-state mixing ratios become identical, $\frac{q_H}{p_H}=\frac{q_L}{p_L}$, so that the heavy and light eigenstates share the same mixing structure. In addition, if direct CP violation in the decay amplitudes is absent, the decay amplitudes of a channel and its CP-conjugate counterpart are related by
\begin{equation}
\bar A_f
=
e^{i\chi_f}A_{\bar f},
\qquad
\bar A_g
=
e^{i\chi_g}A_{\bar g},
\label{eq:CP_decay_relation}
\end{equation}
where \(\chi_f\) and \(\chi_g\) are convention-dependent phases.

Under CP conjugation, the flavor components entering the BI are interchanged according to Eq.~\eqref{eq:H_CP}, while the decay amplitudes are replaced by their CP-conjugate counterparts. Consequently, each overlap appearing in
\(\bar{\Delta}_4^{(\bar f,\bar g)}\)
is obtained from the corresponding overlap in
\(\Delta_4^{(f,g)}\)
by complex conjugation. Since the BI is rephasing invariant, the phases \(\chi_f\) and \(\chi_g\) cancel pairwise in the product of overlaps. One therefore obtains
\begin{equation}
\bar{\Delta}_4^{(\bar f,\bar g)}
=
\left[
\Delta_4^{(f,g)}
\right]^*,
\qquad
(z=0,\ \mathrm{no\ direct\ CP\ violation}),
\label{eq:Delta4_conjugate}
\end{equation}
which follows from the identical mixing structure in the CPT-conserving limit together with the CP-conjugation properties
of the decay amplitudes.

It then follows that
\begin{equation}
\mathcal I_{\mathrm{CPT}}^{(f,g)}
=
\left|
\Delta_4^{(f,g)}
\right|^2,
\qquad
(z=0,\ \mathrm{no\ direct\ CP\ violation}),
\end{equation}
and therefore
\begin{equation}
\mathcal G_{\mathrm{CPT}}^{(f,g)}
=
0,
\qquad
(z=0,\ \mathrm{no\ direct\ CP\ violation}).
\label{eq:G_CPT_zero}
\end{equation}
Equation~\eqref{eq:G_CPT_zero} therefore establishes the geometric observable as a null test of CPT symmetry under the assumptions stated above.

To determine the leading CPT-violating correction, we expand the invariant around the CPT-conserving point \(z=0\). Since the mixing coefficients depend analytically on \(z\) in the neighborhood of the CPT-conserving point \(|z| < 1\), the invariant \(\Delta_4^{(f,g)}\) likewise admits a Taylor expansion about the CPT-conserving point. Writing
\begin{equation}
\Delta_4^{(f,g)}
=
\Delta_4^{(0)}
\left[
1
+
z\,\Xi_{fg}
\right]
+
\mathcal O(z^2),
\label{eq:Delta4_expansion}
\end{equation}
the coefficient
\begin{equation}
\Xi_{fg}
\equiv
\left.
\frac{1}{\Delta_4^{(0)}}
\frac{\partial \Delta_4^{(f,g)}}{\partial z}
\right|_{z=0}
\label{eq:Xi_definition}
\end{equation}
is a dimensionless, channel-dependent quantity that characterizes the sensitivity of the invariant to CPT-violating deformations.

The corresponding expansion of the CP-conjugate invariant is
\begin{equation}
\bar{\Delta}_4^{(\bar f,\bar g)}
=
\left[
\Delta_4^{(0)}
\right]^*
\left[
1
-
z^*\,\Xi_{fg}^*
\right]
+
\mathcal O(z^2).
\label{eq:Delta4bar_expansion}
\end{equation}

Substituting Eqs.~\eqref{eq:Delta4_expansion} and \eqref{eq:Delta4bar_expansion} into Eq.~\eqref{eq:I_CPT} and retaining terms linear in \(z\), we obtain
\begin{align}
\mathcal I_{\mathrm{CPT}}^{(f,g)}
&=
\left|
\Delta_4^{(0)}
\right|^2
\left[
1
+
z\Xi_{fg}
-
z^*\Xi_{fg}^*
\right]
+
\mathcal O(z^2)
\nonumber\\[2mm]
&=
\left|
\Delta_4^{(0)}
\right|^2
\left[
1
+
2i\,\mathrm{Im}
\!\left(
z\,\Xi_{fg}
\right)
\right]
+
\mathcal O(z^2).
\label{eq:I_CPT_linear}
\end{align}

Taking the argument of Eq.~\eqref{eq:I_CPT_linear} yields
\begin{equation}
\mathcal G_{\mathrm{CPT}}^{(f,g)}
=
2\,\mathrm{Im}
\!\left(
z\,\Xi_{fg}
\right)
+
\mathcal O(z^2).
\label{eq:G_CPT_master}
\end{equation}

Equation~\eqref{eq:G_CPT_master} establishes the leading-order relation between the CPT-violating mixing parameter and the geometric observable constructed from the BI. The dependence on CPT violation is encoded through the parameter \(z\), while the channel-dependent
coefficient \(\Xi_{fg}\) quantifies how the CPT-induced deformation is reflected in the selected decay channels. Since \(\Xi_{fg}\) inherits the rephasing invariance of the underlying BI, the observable
\(\mathcal G_{\mathrm{CPT}}^{(f,g)}\) provides a rephasing-invariant geometric probe of CPT violation.

In the following section, we derive the explicit form of \(\Xi_{fg}\) and analyze the conditions under which the resulting geometric observable exhibits maximal sensitivity to CPT-violating effects.

\section{Channel Dependence and Selection Rules for Geometric CPT Sensitivity}

Equation~\eqref{eq:G_CPT_master} shows that the geometric CPT observable is governed not only by the CPT-violating parameter \(z\), but also by the channel-dependent coefficient \(\Xi_{fg}\). The existence of a nonvanishing geometric signal therefore depends on the choice of the decay channels entering the BI. In this section, we derive the explicit form of \(\Xi_{fg}\) directly from the exact invariant and identify the conditions under which the resulting geometric observable is sensitive to CPT violation.

Rather than introducing the conventional quantities
\(
\lambda_f=(q/p)(\bar A_f/A_f)
\)
\cite{Nir2015}, which are no longer uniquely defined once CPT violation leads to the distinct heavy- and light-state mixing ratios,
\(
q_H/p_H \neq q_L/p_L
\),
we work directly with the decay-amplitude ratios
\begin{equation}
\rho_f
\equiv
\frac{\bar A_f}{A_f},
\qquad
\rho_g
\equiv
\frac{\bar A_g}{A_g},
\label{eq:rho_definition}
\end{equation}
which remain well defined independently of the mixing structure.

Starting from the exact expression for the fourth-order BI in Eq.~(\ref{eq:Delta4_explicit}), and substituting $\bar A_f=\rho_f A_f$ and $\bar A_g=\rho_g A_g$, together with their complex conjugates, one obtains
\begin{equation}
\Delta_4^{(f,g)}
=
\frac{
|A_f|^2 |A_g|^2
\,
(q_L+p_L\rho_f^{*})
(p_L-q_L\rho_f)
(q_H-p_H\rho_g^{*})
(p_H+q_H\rho_g)
}{
(p_Hq_L+p_Lq_H)^2
}.
\label{eq:Delta4_rho1}
\end{equation}

Introducing the mixing ratios
\begin{equation}
R_H
\equiv
\frac{q_H}{p_H},
\qquad
R_L
\equiv
\frac{q_L}{p_L},
\label{eq:RHL_definition}
\end{equation}
Eq.~\eqref{eq:Delta4_rho1} may be written as
\begin{equation}
\Delta_4^{(f,g)}
=
|A_f|^2 |A_g|^2
\,
\frac{
(R_L+\rho_f^{*})
(1-R_L\rho_f)
(R_H-\rho_g^{*})
(1+R_H\rho_g)
}{
(R_H+R_L)^2
}.
\label{eq:Delta4_rho_exact}
\end{equation}

Equation~\eqref{eq:Delta4_rho_exact} provides an exact expression for the BI in terms of the decay-channel parameters and the heavy- and light-state mixing ratios.

To determine the CPT sensitivity, we expand the invariant around the CPT-conserving point \(z=0\). Using
\begin{equation}
R_H
=
r
\sqrt{\frac{1-z}{1+z}},
\qquad
R_L
=
r
\sqrt{\frac{1+z}{1-z}},
\label{eq:RH_RL_exact}
\end{equation}
one finds, to first order in \(z\),
\begin{equation}
R_H
=
r(1-z)
+
\mathcal O(z^2),
\qquad
R_L
=
r(1+z)
+
\mathcal O(z^2).
\label{eq:RH_RL_expanded}
\end{equation}

The coefficient \(\Xi_{fg}\) introduced in Eq.~\eqref{eq:Xi_definition} may then be written as
\begin{equation}
\Xi_{fg}
=
\left.
\frac{\partial}{\partial z}
\ln
\Delta_4^{(f,g)}
\right|_{z=0},
\label{eq:Xi_log_definition}
\end{equation}
which is simply the logarithmic derivative of the exact invariant evaluated at \(z=0\).

Taking the logarithm of Eq.~\eqref{eq:Delta4_rho_exact} gives
\begin{align}
\ln\Delta_4^{(f,g)}
&=
\ln(R_L+\rho_f^{*})
+
\ln(1-R_L\rho_f)
\nonumber\\
&\quad
+
\ln(R_H-\rho_g^{*})
+
\ln(1+R_H\rho_g)
\nonumber\\
&\quad
-
2\ln(R_H+R_L)
+
\ln(|A_f|^2|A_g|^2).
\label{eq:logDelta4}
\end{align}

Differentiating with respect to \(z\) and using
\begin{equation}
\left.
\frac{dR_H}{dz}
\right|_{z=0}
=
-r,
\qquad
\left.
\frac{dR_L}{dz}
\right|_{z=0}
=
r,
\end{equation}
yields
\begin{equation}
\Xi_{fg}
=
\frac{r}{r+\rho_f^{*}}
-
\frac{r\rho_f}{1-r\rho_f}
-
\frac{r}{r-\rho_g^{*}}
-
\frac{r\rho_g}{1+r\rho_g},
\label{eq:Xi_intermediate}
\end{equation}
where we have used \(R_H=R_L=r\) at \(z=0\).
Combining terms associated with each decay channel separately, one obtains
\begin{equation}
\Xi_{fg}
=
\frac{
r\left(1-2r\rho_f-|\rho_f|^2\right)
}{
(r+\rho_f^{*})(1-r\rho_f)
}
-
\frac{
r\left(1+2r\rho_g-|\rho_g|^2\right)
}{
(r-\rho_g^{*})(1+r\rho_g)
}.
\label{eq:Xi_general}
\end{equation}

Equation~\eqref{eq:Xi_general} constitutes the exact channel-dependent coefficient governing the linear response of the geometric observable to CPT violation. Since \(\rho_f\), \(\rho_g\), and \(r\) are invariant under flavor rephasings, the quantity \(\Xi_{fg}\) is manifestly rephasing invariant.

Additional insight may be obtained in the limit of negligible direct CP violation. In this case, the decay-amplitude ratios satisfy $|\rho_f|=|\rho_g|=1$, and may therefore be parameterized as
\begin{equation}
\rho_f
=
e^{-i\phi_f},
\qquad
\rho_g
=
e^{-i\phi_g},
\label{eq:rho_phase}
\end{equation}
where \(\phi_f\) and \(\phi_g\) denote the effective weak phases associated with the corresponding decay channels.

Substituting Eq.~\eqref{eq:rho_phase} into Eq.~\eqref{eq:Xi_intermediate} yields
\begin{equation}
\Xi_{fg}
=
\frac{r}{r+e^{i\phi_f}}
-
\frac{re^{-i\phi_f}}
{1-re^{-i\phi_f}}
-
\frac{r}{r-e^{i\phi_g}}
-
\frac{re^{-i\phi_g}}
{1+re^{-i\phi_g}},
\label{eq:Xi_phase_general}
\end{equation}
which is the exact weak-phase expression valid for arbitrary mixing parameter \(r\).

Further simplification is obtained in the limit $r=1$. In particular, \(r=1\) implies $|H_{12}|=|H_{21}|$, a condition satisfied, for example, when T symmetry is conserved in the mixing Hamiltonian. Under this additional assumption, using standard trigonometric identities, Eq.~\eqref{eq:Xi_phase_general} reduces to
\begin{align}
\Xi_{fg}
&=
\left[
\frac{1}{1+e^{i\phi_f}}
-
\frac{e^{-i\phi_f}}
{1-e^{-i\phi_f}}
\right]
-
\left[
\frac{1}{1-e^{i\phi_g}}
+
\frac{e^{-i\phi_g}}
{1+e^{-i\phi_g}}
\right]
\nonumber\\
&=
\left(
1+i\cot\phi_f
\right)
-
\left(
1+i\cot\phi_g
\right)
\nonumber\\
&=
i
\left(
\cot\phi_f
-
\cot\phi_g
\right).
\label{eq:Xi_cotangent}
\end{align}

Substituting Eq.~\eqref{eq:Xi_cotangent} into
Eq.~\eqref{eq:G_CPT_master}, one obtains
\begin{equation}
\mathcal G_{\mathrm{CPT}}^{(f,g)}
=
2\,
\mathrm{Im}
\left[
iz
\left(
\cot\phi_f
-
\cot\phi_g
\right)
\right]
+
\mathcal O(z^2).
\label{eq:G_CPT_phase}
\end{equation}

Since
\(
\cot\phi_f-\cot\phi_g
\)
is purely real,
\begin{equation}
\mathcal G_{\mathrm{CPT}}^{(f,g)}
=
2\,
\mathrm{Re}(z)
\left(
\cot\phi_f
-
\cot\phi_g
\right)
+
\mathcal O(z^2).
\label{eq:G_CPT_realz}
\end{equation}

Equation~\eqref{eq:G_CPT_realz} shows that, to leading order, the geometric CPT observable is sensitive only to the real part of the CPT-violating parameter. Moreover, the sensitivity is governed entirely by the relative weak-phase combination
\(
(\cot\phi_f-\cot\phi_g)
\).

A necessary condition for linear sensitivity is therefore
\begin{equation}
\phi_f
\neq
\phi_g.
\label{eq:selection_rule}
\end{equation}
When the two decay channels probe identical effective weak phases, one finds $\Xi_{fg}=0$, and consequently $\mathcal G_{\mathrm{CPT}}^{(f,g)}=0+\mathcal O(z^2)$, so that the geometric observable loses its linear sensitivity to CPT violation.

The above result provides a channel-selection rule for geometric CPT observables. It establishes a direct connection between the weak-phase structure of the decay channels and the geometric sensitivity of the resulting observable, thereby identifying the class of decay-mode combinations that maximize the linear response to CPT-violating effects.

As a representative illustration of the above selection rule, consider
the neutral \(B_d\)-meson decay channels \cite{CarterSanda1980, BigiSanda1981, PDG2024}
\begin{equation}
B_d^0 \rightarrow J/\psi K_S,
\qquad
B_d^0 \rightarrow D^+D^-.
\end{equation}
These decay modes probe, in general, different combinations of weak amplitudes and therefore provide a natural setting in which the channel-selection criterion derived above may be applied. Within the framework developed above, the corresponding geometric
response is governed by
\begin{equation}
\Xi_{fg}
=
i
\left(
\cot\phi_{J/\psi K_S}
-
\cot\phi_{D^+D^-}
\right),
\end{equation}
in the limit of negligible direct CP violation and \(r=1\). The
channel-selection rule derived in
Eq.~\eqref{eq:selection_rule} then implies that the leading geometric
response vanishes only if the two effective weak phases coincide.
Whenever $\phi_{J/\psi K_S}\neq\phi_{D^+D^-}$, the coefficient \(\Xi_{fg}\) is nonzero, and the geometric CPT
observable acquires linear sensitivity to the CPT-violating parameter
\(z\) through Eq.~\eqref{eq:G_CPT_master}.

The above example illustrates how experimentally well-established decay channels may be incorporated into the present geometric framework
without requiring additional assumptions beyond the effective weak-phase
description. A dedicated phenomenological analysis, including the
determination of the effective weak phases and the expected
experimental sensitivity, lies beyond the scope of the present work and
is left for future investigation.

In the following section, we connect the geometric observable to the Standard-Model Extension and demonstrate how the Lorentz-violating coefficients induce characteristic sidereal modulations in the geometric observable.

\section{Connection to the Standard-Model Extension}
The channel-selection rule derived in the previous section establishes the conditions under which the geometric observable \(\mathcal G_{\mathrm{CPT}}^{(f,g)}\) exhibits linear sensitivity to CPT violation. To connect this framework with experimentally accessible signatures, it is necessary to relate the CPT-violating parameter \(z\) to an underlying dynamical description. A natural setting for such an interpretation is provided by the Standard-Model Extension (SME), which furnishes a general effective field-theoretic framework for parametrizing Lorentz- and CPT-violating effects \cite{Kostelecky1998, Kostelecky2001}.

Within the SME, CPT violation in neutral meson systems originates from flavor-dependent couplings of the valence quarks to fixed background fields. For a neutral meson composed of quarks \(q_1\) and \(\bar q_2\), the relevant CPT-violating coefficient is conventionally written as \cite{Kostelecky2001}
\begin{equation}
\Delta a_\mu
=
r_{q_1}a_\mu^{q_1}
-
r_{q_2}a_\mu^{q_2}.
\label{eq:Deltaa_definition}
\end{equation}
The coefficients \(a_\mu^{q_i}\) are the fundamental Lorentz-violating
parameters of the SME associated with the constituent quark flavors, whereas the combination \(\Delta a_\mu\) represents the effective meson-level coefficient governing CPT violation in neutral-meson mixing.
The factors \(r_{q_i}\) account for nonperturbative hadronic effects arising from the binding of the constituent quarks into the meson.

Following the standard convention adopted in the SME description of
neutral-meson mixing, the leading CPT-violating contribution to the
difference between the diagonal elements of the effective Hamiltonian is
written as \cite{Kostelecky2001}
\begin{equation}
H_{11}-H_{22}
=
\beta^\mu \Delta a_\mu ,
\label{eq:H11H22_SME}
\end{equation}
where
\begin{equation}
\beta^\mu
=
\gamma
\left(
1,\boldsymbol{\beta}
\right)
\label{eq:beta_fourvector}
\end{equation}
is the meson four-velocity.
Equation~\eqref{eq:H11H22_SME} follows the standard normalization
convention adopted in the SME description of neutral-meson mixing. In
alternative conventions, overall normalization factors may be absorbed
into the definition of the effective Hamiltonian or the coefficient
\(\Delta a_\mu\). Throughout the present work, we adopt the convention
of Eq.~\eqref{eq:H11H22_SME}, so that the geometric parameter \(z\)
inherits the same normalization. Combining Eq.~\eqref{eq:H11H22_SME} with the definition of \(z\) in Eq.~(\ref{eq:z_definition}), one obtains
\begin{equation}
z
=
\frac{\beta^\mu \Delta a_\mu}
{\Delta\lambda}.
\label{eq:z_SME}
\end{equation}

Equation~\eqref{eq:z_SME} provides the dynamical interpretation of the geometric deformation parameter. Within the geometric framework developed in the preceding sections, \(z\) parametrizes the relative deformation between the heavy- and light-state mixing directions, while within the SME it acquires the dynamical interpretation given by Eq.~\eqref{eq:z_SME}.

Substituting Eq.~\eqref{eq:z_SME} into the geometric CPT observable of Eq.~\eqref{eq:G_CPT_master}, we obtain
\begin{equation}
\mathcal G_{\mathrm{CPT}}^{(f,g)}
=
2\,\mathrm{Im}
\left[
\Xi_{fg}
\frac{\beta^\mu \Delta a_\mu}
{\Delta\lambda}
\right]
+
\mathcal O(\Delta a_\mu^2).
\label{eq:G_CPT_SME}
\end{equation}

Equation~\eqref{eq:G_CPT_SME} provides the principal phenomenological connection between the geometric observable constructed from BIs and the Lorentz-violating SME coefficients. It shows that the observable factorizes naturally into a dynamical component $\frac{\beta^\mu \Delta a_\mu}{\Delta\lambda}$, which contains the underlying CPT-violating physics, and a geometric component $\Xi_{fg}$, which depends only on the selected decay channels and determines how efficiently the projective-space deformation is converted into an observable phase.

A particularly transparent expression is obtained in the weak-phase limit discussed in the previous section.
Using the weak-phase result of Eq.~(\ref{eq:Xi_cotangent}), valid in
the combined limit of negligible direct CP violation and
\(r=1\), Eq.~\eqref{eq:G_CPT_SME} reduces to

\begin{equation}
\mathcal G_{\mathrm{CPT}}^{(f,g)}
=
2
\left(
\cot\phi_f
-
\cot\phi_g
\right)
\,
\mathrm{Re}
\left(
\frac{\beta^\mu \Delta a_\mu}
{\Delta\lambda}
\right)
+
\mathcal O(\Delta a_\mu^2).
\label{eq:G_CPT_SME_weakphase}
\end{equation}

Equation~\eqref{eq:G_CPT_SME_weakphase} demonstrates that, at leading order, the geometric CPT observable is sensitive only to the real part of the SME-induced deformation parameter \(z\), equivalently to the real part of \(\left(
\frac{\beta^\mu \Delta a_\mu}
{\Delta\lambda}
\right)\). The decay channels determine the overall sensitivity through the weak-phase combination \((\cot\phi_f-\cot\phi_g)\), while the Lorentz-violating physics enters through the projected SME coefficient \(\beta^\mu\Delta a_\mu\).

Several notable features follow immediately. First, the observable is manifestly rephasing invariant, since both \(z\) and \(\Xi_{fg}\) are independent of flavor conventions. Second, the explicit appearance of the meson four-velocity implies that the geometric observable depends not only on the meson system itself but also on its motion relative to the preferred SME background. Third, the factorized structure of Eq.~\eqref{eq:G_CPT_SME} cleanly separates the geometric response of the decay channels from the underlying Lorentz-violating dynamics.

The velocity dependence encoded in Eq.~\eqref{eq:z_SME} further implies that the geometric CPT observable exhibits characteristic temporal variations as the laboratory frame rotates with respect to the approximately inertial Sun-centered frame conventionally employed in SME analyses. The resulting observable is therefore sensitive not only to the magnitude of CPT violation but also to its
directional dependence through the SME coefficients.

In the following section, we derive the resulting sidereal modulation of the geometric observable and show how the harmonic structure of \(\mathcal G_{\mathrm{CPT}}^{(f,g)}\) directly reflects the underlying SME coefficients.
\section{Sidereal Modulation as a Geometric Signature of Lorentz Violation}

Equation~\eqref{eq:G_CPT_SME} establishes that the geometric CPT observable depends explicitly on the contraction \(\beta^\mu\Delta a_\mu\) between the meson four-velocity and the Lorentz-violating SME coefficient. Since the SME coefficients are conventionally defined in a Sun-centered inertial frame, whereas experiments are performed in laboratories fixed to the rotating Earth, the geometric observable acquires a characteristic sidereal time dependence \cite{Kostelecky1999, KosteleckyRussell2011}. This temporal modulation provides a distinctive signature of Lorentz violation and directly connects the geometric observable to the underlying SME coefficients.

Following the standard SME convention, we adopt the Sun-centered celestial-equatorial frame \((T,X,Y,Z)\), in which the coefficients \(\Delta a_\mu\) are taken to be constant. The \(Z\)-axis is chosen parallel to the Earth's rotation axis, while the \(X\)-axis points from the Earth toward the vernal equinox at the reference epoch \cite{KosteleckyRussell2011}. In this frame, the Lorentz-violating coefficient may be written as
\begin{equation}
\Delta a^\mu
=
\left(
\Delta a_T,
\Delta a_X,
\Delta a_Y,
\Delta a_Z
\right).
\label{eq:Deltaa_sun}
\end{equation}

The CPT-violating parameter entering the geometric framework is $z=\frac{\beta^\mu\Delta a_\mu}{\Delta\lambda}$, where the four-velocity \(\beta^\mu\) is naturally defined in the laboratory frame. As the Earth rotates with sidereal frequency \(\Omega_\oplus\), the orientation of the laboratory frame changes with respect to the Sun-centered frame, producing a time-dependent projection of the spatial SME coefficients onto the meson momentum direction.

Following the standard SME treatment of sidereal variations in neutral meson systems, the contraction $\beta^\mu\Delta a_\mu$ may be written in
the generic first-harmonic form \cite{Kostelecky1999}
\begin{equation}
\beta^\mu\Delta a_\mu
=
\gamma
\left[
\mathcal A_0
+
\mathcal A_c
\cos\!\left(
\Omega_\oplus t_{\mathrm{sid}}
\right)
+
\mathcal A_s
\sin\!\left(
\Omega_\oplus t_{\mathrm{sid}}
\right)
\right],
\label{eq:betaa_sidereal}
\end{equation}
where \(t_{\mathrm{sid}}\) denotes the local sidereal time. The coefficients \(\mathcal A_0\), \(\mathcal A_c\), and \(\mathcal A_s\) are linear combinations of the SME coefficients \(\Delta a_\mu\) and depend on the laboratory colatitude, the beam orientation, and the meson boost distribution. Explicit expressions may be obtained by transforming the meson momentum from the laboratory frame to the Sun-centered frame following the standard SME formalism.

Substituting Eq.~\eqref{eq:betaa_sidereal} into Eq.~\eqref{eq:z_SME} yields
\begin{equation}
z(t_{\mathrm{sid}})
=
\frac{\gamma}{\Delta\lambda}
\left[
\mathcal A_0
+
\mathcal A_c
\cos\!\left(
\Omega_\oplus t_{\mathrm{sid}}
\right)
+
\mathcal A_s
\sin\!\left(
\Omega_\oplus t_{\mathrm{sid}}
\right)
\right].
\label{eq:z_sidereal}
\end{equation}

Equation~\eqref{eq:z_sidereal} shows that the CPT-violating parameter acquires a characteristic sidereal modulation. Geometrically, this corresponds to a periodic variation of the projective-space deformation parameter \(z\) arising from the changing orientation of the laboratory frame relative to the preferred SME background as the Earth rotates.

The geometric CPT observable therefore inherits the same harmonic structure. Using Eq.~\eqref{eq:G_CPT_master}, one finds
\begin{align}
\mathcal G_{\mathrm{CPT}}^{(f,g)}
(t_{\mathrm{sid}})
&=
2\,
\mathrm{Im}
\!\left[
\Xi_{fg}
\,z(t_{\mathrm{sid}})
\right]
+
\mathcal O(z^2)
\nonumber\\[2mm]
&=
\mathcal G_0
+
\mathcal G_c
\cos\!\left(
\Omega_\oplus t_{\mathrm{sid}}
\right)
+
\mathcal G_s
\sin\!\left(
\Omega_\oplus t_{\mathrm{sid}}
\right),
\label{eq:G_CPT_sidereal}
\end{align}
where $\mathcal G_0=2\,\mathrm{Im}\!\left(\Xi_{fg}\frac{\gamma\mathcal A_0}{\Delta\lambda}\right)$, $\mathcal G_c=2\,\mathrm{Im}\!\left(\Xi_{fg}\frac{\gamma\mathcal A_c}{\Delta\lambda}\right)$, and $\mathcal G_s=2\,\mathrm{Im}\!\left(\Xi_{fg}\frac{\gamma\mathcal A_s}{\Delta\lambda}\right)$.

A particularly transparent form is obtained in the weak-phase limit discussed in the previous section. Using Eq.~(\ref{eq:Xi_cotangent}), we have, in this limit,
\begin{equation}
\mathcal G_{\mathrm{CPT}}^{(f,g)}
(t_{\mathrm{sid}})
=
2
\left(
\cot\phi_f
-
\cot\phi_g
\right)
\,
\mathrm{Re}
\!\left[
z(t_{\mathrm{sid}})
\right]
+
\mathcal O(z^2),
\label{eq:G_CPT_weakphase_sidereal}
\end{equation}
or equivalently,
\begin{align}
\mathcal G_{\mathrm{CPT}}^{(f,g)}
(t_{\mathrm{sid}})
&=
\widetilde{\mathcal G}_0
+
\widetilde{\mathcal G}_c
\cos\!\left(
\Omega_\oplus t_{\mathrm{sid}}
\right)
+
\widetilde{\mathcal G}_s
\sin\!\left(
\Omega_\oplus t_{\mathrm{sid}}
\right),
\end{align}
with
\begin{equation}
\widetilde{\mathcal G}_i
=
2
\left(
\cot\phi_f
-
\cot\phi_g
\right)
\,
\mathrm{Re}
\!\left(
\frac{\gamma\mathcal A_i}
{\Delta\lambda}
\right),
\qquad
i=0,c,s.
\label{eq:Gtilde_coeffs}
\end{equation}

Equation~\eqref{eq:G_CPT_weakphase_sidereal} demonstrates that, at leading order, the sidereal modulation probes the real part of the SME-induced deformation parameter. The weak-phase combination \((\cot\phi_f-\cot\phi_g)\) controls the geometric sensitivity, while the harmonic coefficients encode the underlying Lorentz-violating physics.

Several features of this result merit emphasis. First, the sidereal modulation arises from the time-dependent projection of the meson four-velocity onto the fixed SME background, which is encoded in the velocity dependence of the CPT-violating parameter \(z\). Second, the observable remains manifestly rephasing invariant throughout the construction. Third, the channel-selection rule derived in the previous section implies that only decay-channel pairs with distinct effective weak phases contribute to the harmonic amplitudes.

The emergence of sidereal modulation provides a clear phenomenological signature of the geometric framework developed in this work. Complementing conventional decay-rate asymmetries, the present approach identifies Lorentz-violating effects through a rephasing-invariant geometric observable whose temporal structure is determined by the Earth's rotation relative to the preferred SME background.

A detailed experimental analysis of the sensitivity to the coefficients \(\Delta a_\mu\) lies beyond the scope of the present work. Nevertheless, Eq.~\eqref{eq:G_CPT_sidereal} provides an explicit leading-order relation between measurable sidereal variations of the geometric observable and the underlying Lorentz-violating parameters, thereby providing a geometric framework for future searches for CPT and Lorentz violation in neutral meson systems.

\section{Summary}
In this work, we have developed a geometric formulation for describing CPT violation in neutral meson systems based on Bargmann invariants (BIs) within the biorthogonal framework adopted throughout this work. Starting from the effective Hamiltonian governing neutral meson mixing, we interpreted CPT violation as a geometric deformation of the heavy- and light-state mixing structure and constructed a rephasing-invariant fourth-order BI from the heavy- and light-mass eigenstates together with experimentally accessible decay channels.

Building upon this construction, we introduced a geometric CPT observable defined through the phase of a product of the fourth-order BI and its CP-conjugate counterpart, and showed that its leading response is governed by the logarithmic variation of the invariant with respect to the CPT-violating mixing parameter. This formulation naturally yields a channel-dependent geometric coefficient that quantifies the sensitivity of different decay
channels to CPT-violating effects and leads to a simple
channel-selection criterion identifying decay-mode combinations capable of exhibiting linear geometric sensitivity.

We further established the connection between the geometric framework and the Standard-Model Extension by relating the geometric deformation parameter to the Lorentz-violating SME coefficients. Within this
correspondence, the geometric observable inherits the characteristic sidereal modulation arising from the Earth's rotation, thereby providing a geometric interpretation of the directional dependence of
Lorentz-violating effects in neutral meson mixing. In this way, the present formalism connects the geometric structure encoded in BIs with the established phenomenology of CPT and Lorentz violation.

The geometric framework developed in this work provides a rephasing-invariant characterization of CPT violation that naturally links the structure of neutral-meson mixing with experimentally accessible decay channels and the Lorentz-violating phenomenology of the Standard-Model Extension. Beyond offering a complementary geometric perspective on unstable quantum systems, the framework establishes a systematic foundation for future theoretical and phenomenological studies of geometric signatures of CPT and Lorentz violation in neutral meson systems, including analyses of specific decay channels, correlated meson pairs, and related non-Hermitian quantum systems.

{\it Acknowledgement :} I wish to thank Prof. Utpal Sarkar and Prof. Arghya Taraphder for support and encouragement. I would like to thank MoE, Government of India for the research fellowship.

\end{document}